\documentclass[useAMS,usenatbib,usegraphicx]{mn2e}
\usepackage{url}

\begin{document}

\title[The GMRT-EoR Experiment: H\,{\normalsize \it I} Power Spectrum]
 {The GMRT Epoch of Reionization Experiment: A new upper limit on the neutral hydrogen power spectrum at $\bmath{z\approx8.6}$}

\author[G. Paciga et al.]
 {Gregory~Paciga$^1$\thanks{Email:paciga@astro.utoronto.ca},
   Tzu-Ching~Chang$^{1,2}$,
   Yashwant~Gupta$^3$,
   Rajaram~Nityanada$^3$,
   \newauthor
   Julia~Odegova$^1$,
   Ue-Li~Pen$^1$\thanks{Email:pen@cita.utoronto.ca},
   Jeffrey~B.~Peterson$^4$,
   Jayanta~Roy$^3$,
   Kris~Sigurdson$^5$\\
   $^1$ CITA, University of Toronto, 60 St George Street, Toronto, ON M5S 3H8, Canada \\
   $^2$ IAA, Academia Sinica, P.O. Box 23-141, Taipei 10617, Taiwan \\
   $^3$ National Center for Radio Astrophysics, Tata Institute for Fundamental Research, Pune 411 007, India \\
   $^4$ Department of Physics, Carnegie Mellon University, 5000 Forbes Ave, Pittsburgh, PA 15213, USA \\
   $^5$ Department of Physics and Astronomy, University of British Columbia, Vancouver, BC V6T 1Z1, Canada \\
}

\maketitle

\begin{abstract}
We present a new upper limit to the 21\,cm power spectrum during
the Epoch of Reionization (EoR)
which constrains reionization models with an unheated IGM.
The GMRT-EoR experiment is an ongoing effort
to make a statistical detection of the power spectrum of
21\,cm neutral hydrogen emission at redshift $z\sim9$.
Data from this redshift constrain models of the EoR,
the end of the Dark Ages arising from the formation of the first bright UV sources,
probably stars or mini-quasars. 
We present results from approximately 50 hours of observations
at the Giant Metrewave Radio Telescope in India from December 2007.
We describe radio frequency interference
(RFI) localisation schemes which allow bright sources on
the ground to be identified and physically removed
in addition to automated flagging.
Singular-value decomposition is used to remove remaining broadband RFI
by identifying ground sources with large eigenvalues.
Foregrounds are modelled using a piecewise linear filter
and the power spectrum is measured using cross-correlations 
of foreground subtracted images.
\end{abstract}


%%%%%%%%%%%%%%%%%%%%%%%%%%%%%%%%%%%%%%%%%%%%%%%%%%%%%%%%%%%%%%%%%%%%%%%%%%%%%%%%
\section{Introduction}
\label{sec:Introduction}
%%%%%%%%%%%%%%%%%%%%%%%%%%%%%%%%%%%%%%%%%%%%%%%%%%%%%%%%%%%%%%%%%%%%%%%%%%%%%%%%

Between the recombination epoch at $z \sim 1100$
and the first round of star formation at $z \sim 10$
the Universe was filled with neutral hydrogen.
This neutral gas is thought to have produced 21\,cm hyperfine emission
with an effective continuum brightness temperature
between -500 and 30\,mK \citep{Furlanetto06}
using $z=8.6$ and \emph{WMAP} 7-year cosmological parameters \citep{Komatsu10},
which have formal error bars of roughly 10 per cent. 
As the first stars formed,
beginning at local peaks in the matter density,
the hydrogen gas was locally ionized.
This is the  start of a period of cosmic evolution called
the Epoch (or Era) of Reionization (EoR).
Over time the ionized cells grew and overlapped,
creating a patchwork of ionized and neutral cells.
This general patchy topology is well motivated by theory and simulations
(e.g. \citealt{Furlanetto04,Iliev06,McQuinn07,Zahn07}; see \citealt{Trac09} for a review)
though the exact properties are poorly constrained.
Eventually, by $z \sim 6$,
the ionization of the broadly distributed material was complete,
leaving only rare pockets of neutral gas.

In addition to directly constraining 
the redshift of last scattering
$z_\mathrm{ls} \sim 1100$,  
Cosmic Microwave Background (CMB) polarization data have been used to constrain 
the redshift to the range $8.0<z<12.8$
\citep[WMAP7;][]{Komatsu10} under the assumption it was instantaneous,
although evidence from the CMB indicates that reionization was an extended process
\citep{Dunkley09}.
There is also substantial information on the neutral content of the
intergalactic medium (IGM) since $z \sim 6$
via observations of the Gunn-Peterson trough \citep{Gunn65}
in quasar spectra \citep{Djorgovski01, Becker01, Fan06, Willott07}.
Lyman-alpha (Ly-$\alpha$) absorption provides a sensitive probe of neutral gas density and
can be used to constrain reionization \citep[e.g.,][]{Fan02}.
With these two techniques we can observe the start 
and the end of the neutral era.
However, neither CMB nor quasar observations allow detailed examination 
of the reionization era itself.
For this it has long been proposed to use 21\,cm fluctuations
in the brightness temperature of neutral hydrogen
\citep{Sunyaev75, Hogan79}.
The 21\,cm power spectrum, resulting from a combination
of the patchy ionized and neutral medium and the underlying
mass power spectrum,
is generally considered one of the
most promising signals \citep{Scott90, Zaldarriaga04}
and much attention has been paid in the literature toward
designing suitable experiments to detect it
\citep[e.g.,][]{Morales04, Morales05, Bowman06, Harker10}.

Here the observing frequency is 1420\,MHz$(1+z)^{-1}$, 
in the very high frequency (VHF) range of the radio spectrum near 150\,MHz.
Several programs are underway to study the EoR in the VHF band
including \mbox{LOFAR}\footnote{\url{http://www.lofar.org/}}
(\citealt{Kassim04, Rottgering06}; for the EoR case see, e.g., \citealt{Zaroubi05, Harker10}),
\mbox{MWA}\footnote{\url{http://www.mwatelescope.org/}}
(\citealt{Lonsdale09}; for the EoR case see, e.g., \citealt{Morales06MWA, Bowman06, Lidz08}),
\mbox{PAPER}\footnote{\url{http://astro.berkeley.edu/~dbacker/eor/}} \citep{Parsons10},
\mbox{21CMA}\footnote{\url{http://web.phys.cmu.edu/~past/}} \citep[a.k.a. PaST;][]{Peterson04},
in addition to the current \mbox{GMRT}\footnote{\url{http://gmrt.ncra.tifr.res.in/}} program,
and the first goal of such efforts is to
determine the redshift at which roughly half the volume of the universe is ionized. 
Because the 21\,cm emission is an isolated single line, 
these observations can provide three dimensional information 
over a large redshift range \citep{Loeb04, Furlanetto06}, 
allowing a broad search for the half-ionization redshift.
In the future 21\,cm tomography has the potential
to add strong constraints on cosmological parameters
\citep[e.g.,][]{McQuinn06,Cooray08,Mao08,Furlanetto09,Masui10}.

The brightness temperature of the 21\,cm line relative to the
CMB is determined by the ionization fraction of hydrogen and
the spin temperature of the neutral population,
which is in turn governed by the background radiation and the
kinetic temperature of the gas \citep{Purcell56, Field59, Furlanetto06}.
Reionization requires a minimum expenditure of 13.6\,eV of energy per hydrogen atom.
In contrast, Ly-$\alpha$ photons, 
when absorbed by a neutral atom, are quickly re-emitted. 
Ly-$\alpha$ photons are said to undergo `resonant scattering' 
and each (multiply-scattered) photon can affect many atoms 
before cosmic redshifting makes them ineffective
\citep{Wouthuysen52, Field59, Chuzhoy06}.
This means Ly-$\alpha$ pumping of the hyperfine transition 
requires only about 1 per cent of the UV flux required for ionization. 
Assuming a gradual increase in UV flux with time, 
well before flux levels for reionization are reached, 
Ly-$\alpha$ pumping will couple the spin temperature 
to the kinetic temperature of the gas.  
If there was no source of heat at that era other than a weak UV flux, 
the gas kinetic temperature must have been at its adiabatic expansion value 
1.7\,K at $z=8.5$,
and neutral gas will produce a signal due to absorption of CMB photons 
in excess of stimulated emission 
and ionized structures would be seen as low-brightness regions on the sky \citep{Chen04}. 
In such a cold-gas model 
the brightness temperature of the neutral gas 
against the CMB
can be as low as -500\,mK.

A small fraction of the mass will have collapsed into minihaloes, which
have a temperature higher than this adiabatic temperature
\citep{Shapiro06}.  This fraction may even be substantially smaller
due to non-perturbative velocity flows \citep{Tseliakhovich10, Dalal10}.
No study was found on the impact of these non-linear effects on
the Ly-$\alpha$ pumped IGM temperature.  At linear order, the
positive and negative over and under densities cancel, and the non-linear
collapse fraction is still small, so we expect the realistic value to
be similar to the adiabatic prediction.

Alternatively, 
X-rays from supernovae or QSOs might have heated the IGM 
above the CMB temperature of $\approx$\,30\,K before reionization
\citep{Madau97, Chen04}. 
X-ray heating between 30\,K and 10\,000\,K 
will result in a largely neutral but warm IGM, 
which would be seen in emission.  
In the limit that $T_\mathrm{s} \gg 30$\,K, 
the volume emissivity becomes independent of temperature.
Patchy X-ray heating can also result in large angular scale structure \citep{Alvarez10}.
The sky brightness temperature of the neutral gas 
has an asymptote at $\sim$\,30\,mK \citep{Furlanetto06}.

The cosmic luminosity of X-rays at $z\sim9$ is not known
and difficult to estimate \citep[see e.g.,][]{Dijkstra04, Salvaterra07, Guo09}.
If the rate of core collapse SNe at high redshift matches that of today, 
the X-ray output at $z \sim 9$ would have been sufficient 
to raise the IGM temperature above the CMB temperature at the onset of reionization. 
However, the mechanism of and factors affecting core-collapse are not known 
and most numerical models appear 
to generically not result in SNe at all \citep{Mezzacappa05}.  
It is possible that at high $z$, 
core collapse SNe were not as abundant as today, 
and the IGM was still in absorption during the EoR. 
We therefore consider two limits: 
one where the IGM is still cold and $T_\mathrm{b} = -500$\,mK, 
and another where it is heated above CMB and $T_\mathrm{b} = 30$\,mK.
A general parametrization of these scenarios was recently proposed
by \citet{Pritchard10}.

The GMRT-EoR project uses 
the Giant Metrewave Radio Telescope \citep[GMRT;][]{GMRT} near Pune, India, 
to make measurements of the power spectrum of the neutral hydrogen signal 
with the hope of characterizing the structure in the range $8.1 < z < 9.2$. 
The FWHM of the GMRT primary beam at 150\,MHz is 3.3$^\circ$ 
which provides a cylindrical comoving survey volume of $(280$\,$h^{-1}$Mpc$)^3$,
with about equal dimensions in three directions. 
The primary sensitivity comes from the compact central core 
which is contained within $\approx$\,1\,km, or 30 dish diametres,
which gives images with $\sim 30^2$ resolution elements.
As recorded, 
the data have high spectral resolution along the line of sight,
but we bin the data for comparable transverse and radial resolution.
Each data cube has $\sim 30^3 \approx 30000$ resolution elements 
so even though the signal to noise ratio of each element is less than one, 
the spatial power can be measured to 1 per cent.  
A statistical measurement of the power spectrum 
provides the best hope of pinning down the EoR
half-ionization redshift.

In section~\ref{sec:Observations} we describe the data
and analysis, including RFI removal and foreground filters used.
In section~\ref{sec:Multiday} we extend the analysis to
cross-correlations of multiple nights, 
and a measurement of the power spectrum.
We then conclude in section~\ref{sec:Conclusion}.
All distances are in comoving coordinates.

%%%%%%%%%%%%%%%%%%%%%%%%%%%%%%%%%%%%%%%%%%%%%%%%%%%%%%%%%%%%%%%%%%%%%%%%%%%
\section{Observations}
\label{sec:Observations}
%%%%%%%%%%%%%%%%%%%%%%%%%%%%%%%%%%%%%%%%%%%%%%%%%%%%%%%%%%%%%%%%%%%%%%%%%%%

\subsection{GMRT and Data Description}
\label{sec:GMRT and GSB}

The GMRT is a radio interferometer consisting of 30 antennas, 
each with a diametre of 45\,m. 
Fourteen of these are arranged in a dense central core within 1\,km 
which allows the high brightness sensitivity required 
to search for the dim EoR signal \citep{Pen09}. 
The longest separation between antennas is about 25\,km.  
For this experiment the telescope was operated at 150\,MHz 
with a 16.7\,MHz bandwidth. 
For 21\,cm (1421\,MHz) emission, 
this frequency range probes a redshift range of $8.1 < z < 9.2$. 
The highest angular resolution at this observing frequency at GMRT 
is about 20 arcseconds.

Each antenna provides a pair of left and right circularly polarized signals, 
which are passed through a new signal processing system, 
built in part for this project,
called the GMRT Software Backend \citep[GSB;][]{Roy10}.  
The final recorded visibilities have a resolution of 
7.8\,kHz and 1/4\,s
in frequency and time, respectively.

\begin{figure}
  \begin{center}
    \includegraphics[width=\columnwidth]{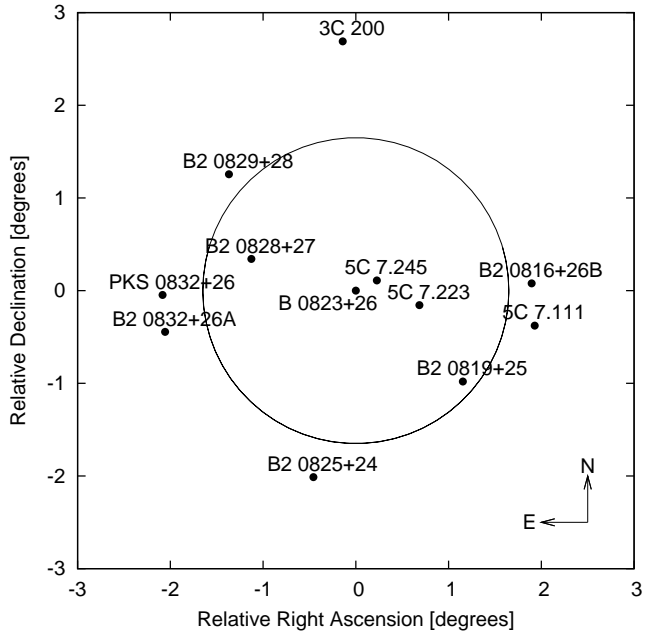}
  \end{center}
  \caption{The 12 brightest sources in the field used for these observations.
    Pulsar PSR~B0823+26 is at the centre of the field.
    The circle denotes the half-power diametre of the main beam. }
  \label{fig:sources}
\end{figure}

The field is centred on the pulsar B0823+26. 
This pulsar has a period of 0.53 seconds 
and an average flux of 350\,mJy at 150\,MHz \citep{Hobbs04}.
It is situated in a relatively cold part of the sky 
at galactic latitude $\approx 30^\circ$
with few nearby bright sources. 
The on-pulse flux is about 6\,Jy, 
brighter than all other sources in the field, 
making it a good calibrator.
The calibration of this data was described in \citet{Pen09} and is summarized below.
The positions of the brightest sources in the field are shown in Fig.~\ref{fig:sources}.

The data are folded in time into 16 `gates' 
such that the pulse from the pulsar is contained within one gate. 
The period of the pulsar is much shorter than 
the time-scale at which ionospheric fluctuations dominate, 
so by comparing the `on' gate with the neighbouring `off' gates, 
everything in the field 
that is constant over the period of the pulsar can be removed. 
This includes sky sources 
and most radio frequency interference (RFI).  
By comparing the pulsar signal received by each antenna, 
the relative system gain of each antenna can be calibrated.  
This technique allows calibration of both phase and polarization in real time, 
with 100 per cent observing efficiency, 
since no additional time is needed for phase or flux calibration.  
Errors in clock synchronization sometimes cause 
the pulsar pulse to straddle two gates. 
In these cases it is necessary to include the flanking gates 
when identifying the pulsar signal. 
This is not ideal 
since the pulsar signal is diluted over multiple gates, 
but the correction is limited to 
only the portion of observations that require it.

% Noise injection and radio source
Since the pulsar amplitude varies from pulse to pulse, 
the absolute system gain needs to be measured separately. 
This is done using a noise injection system, 
which the GMRT software backend decodes 
to calibrate the absolute gain of the system, 
while a sky radio source is used 
to transfer the noise source calibration on to the sky. 
The primary flux calibrator used was 5C~7.245, 
a radio galaxy at $z\approx1.6$ \citep{Willott01},
located within the field of view (see Fig.~\ref{fig:sources}).
However, since this radio galaxy is an extended object, 
it can only be used as a calibrator for short baselines 
where its structure is not resolved. 
For antennas in longer baselines 
we determine the relative gain calibration using the pulsar.

\subsection{Data Analysis}
\label{sec:Data}

Data from 2007 December 10, 11, 14, 16, 17, and 18 
are included in the current analysis.  
A software pipeline has been developed 
to automate the calibration and interference removal steps 
described below. 
The current configuration takes approximately 11 hours 
to process one hour of data on the CITA Sunnyvale computing cluster, 
or 20 hours on the National Centre for Radio Astronomy (NCRA) HP computing cluster 
in Pune, India. 
The NCRA facility can process four hours in parallel, 
completing one night of observations in slightly less than two days. 
While the Sunnyvale cluster has the capacity 
to process as much as 12 hours of data at once, 
storage capacity, shared usage, and other bottlenecks 
reduce this significantly.

One of the limiting factors for measurement of the EoR signal 
is broadband radio frequency interference (RFI), 
which dominates the signal at 150\,MHz. 
In addition to the standard procedure of flagging bad data,
we also use a singular value decomposition (SVD)
to remove broadband RFI from the data,
and also to identify and remove interference sources.
We believe this approach is unique among other
RFI mitigation strategies in the radio community.

Narrow line interference was removed 
by masking points in each frequency bin 
with an intensity above some threshold. 
This was done twice in the data reduction pipeline. 
Input data are initially flagged with a threshold of 
$8\sigma$ on a Gaussian scale, 
and then again at $3\sigma$ 
after broadband interference was removed. 
The first mask can not be too aggressive 
or the techniques to remove broadband interference, 
discussed below, will fail.

\begin{figure}
  \begin{center}
    \includegraphics[width=\columnwidth]{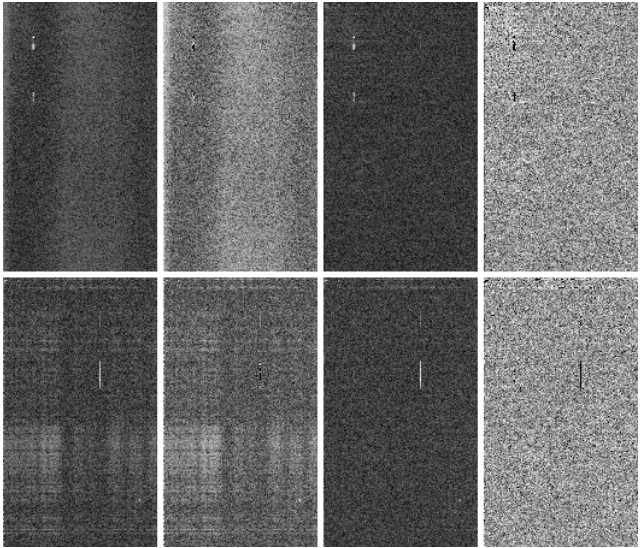}
  \end{center}
  \caption{Representative stages in the data reduction pipeline.
    In each panel, the horizontal axis is frequency,
    covering approximately 1\,MHz,
    and the vertical is time, increasing downward, 
    and covering approximately one hour. 
    The grey-scale is the cross-power spectral density. 
    The top row is the C0-C8 baseline 
    and the bottom row the C0-W4 baseline, 
    which are approximately 560\,m and 9400\,m respectively.  
    Large bright patches indicate broadband RFI, 
    and vertical lines indicate line RFI. 
    The first panel in each row is the initial input data. 
    The second is after the initial $8\sigma$ mask, 
    with most interference still visible. 
    The third panel is after removing the largest eigenvalues in the SVD. 
    The broadband interference is no longer visible. 
    The last panel shows the final $3\sigma$ mask removing the line RFI, 
    leaving a nearly uniform image.}
  \label{fig:RFI-removal}
\end{figure}

\begin{figure}
  \begin{center}
    \includegraphics[width=\columnwidth]{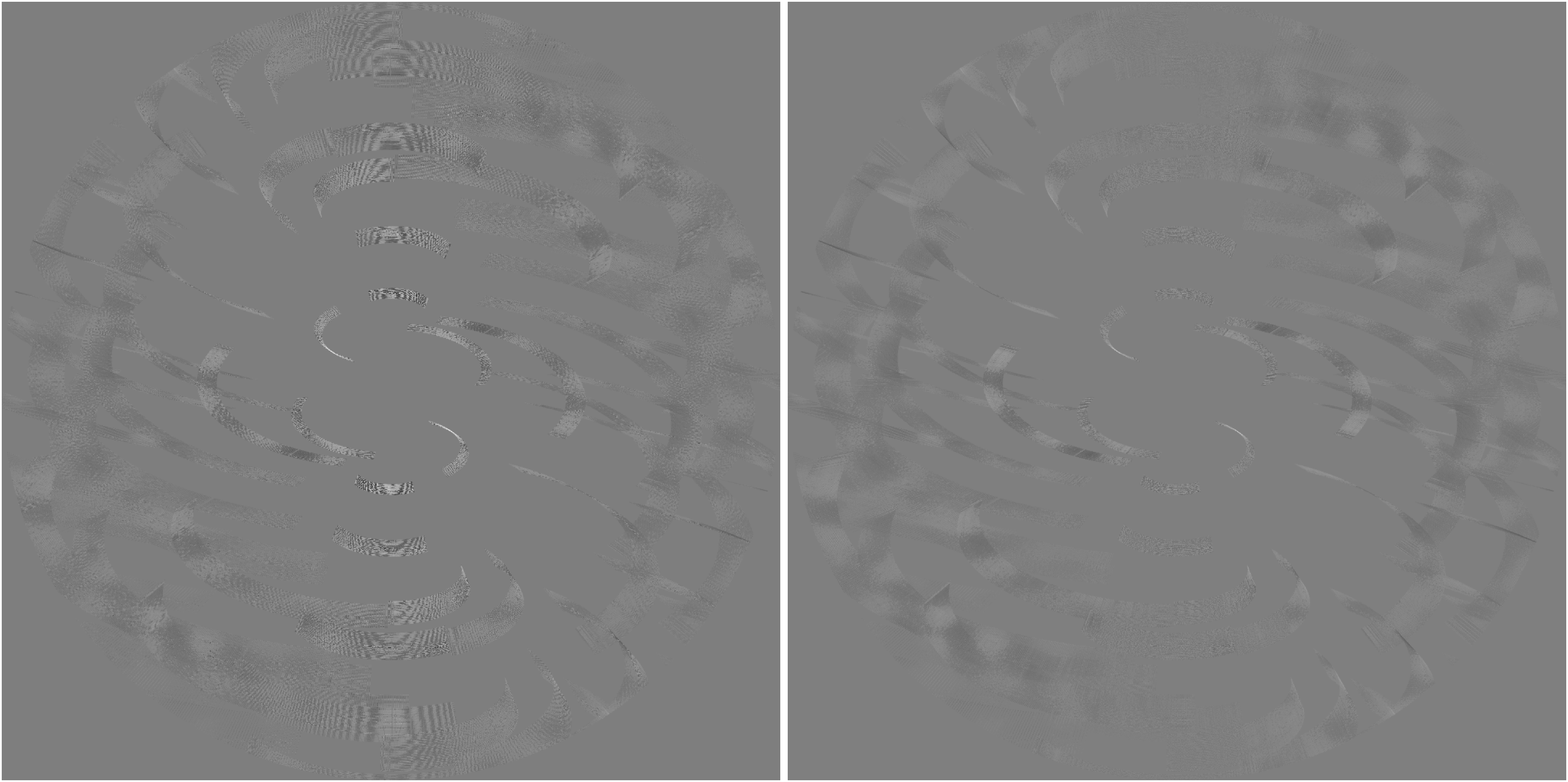}
  \end{center}
  \caption{Raw visibilities with $(u,v)$ distance $|{\bf b}|<200$
    before the SVD RFI removal step (left)
    and after the RFI removal (right).
    The bright patterns in the central vertical strip (low $|u|$) are caused by RFI,
    and are significantly reduced by the SVD procedure.
    The effect of this process on the final power spectrum is considered
    in section~\ref{sec:cc-and-ps}.}
  \label{fig:before-and-after-vis}
\end{figure}

% RFIO REMOVAL (SVD)
Singular value decomposition (SVD) is used 
to separate broadband radio sources on the ground from those in the sky. 
Ground-based sources contribute most 
to the largest eigenvalues 
since they do not move as a function of time with respect to the array, 
while sky sources rotate.  
One hour of data has 14396 time records, 
and each record has about 7.5 million entries 
corresponding to the number of frequency channels 
and baselines between the 60 antennas. 
This is treated as a matrix with each time record as one row. 
The 50 largest eigenvalues are identified 
through an SVD and flagged as noise to be removed.
A sample of the data at a few intermediate stages 
showing the successful removal of both line and broadband RFI 
can be seen in Fig.~\ref{fig:RFI-removal}. 
The RFI patterns in $(u,v)$ space 
both before and after the SVD
are illustrated in Fig.~\ref{fig:before-and-after-vis}.

\subsection{Physical Removal of Radio Interference Sources}
\label{sec:Physical Radio Interference Removal}

While some RFI can be removed in the data analysis,
this risks removing sky signal as well. 
Ideally one would prevent RFI from occurring at all 
by identifying and correcting the physical sources.  
We can take advantage of the fact that as an interferometer 
GMRT is able to make images of the near-field, 
and create maps of bright RFI sources near the GMRT itself.

\begin{figure}
  \begin{center}
    \includegraphics[width=\columnwidth]{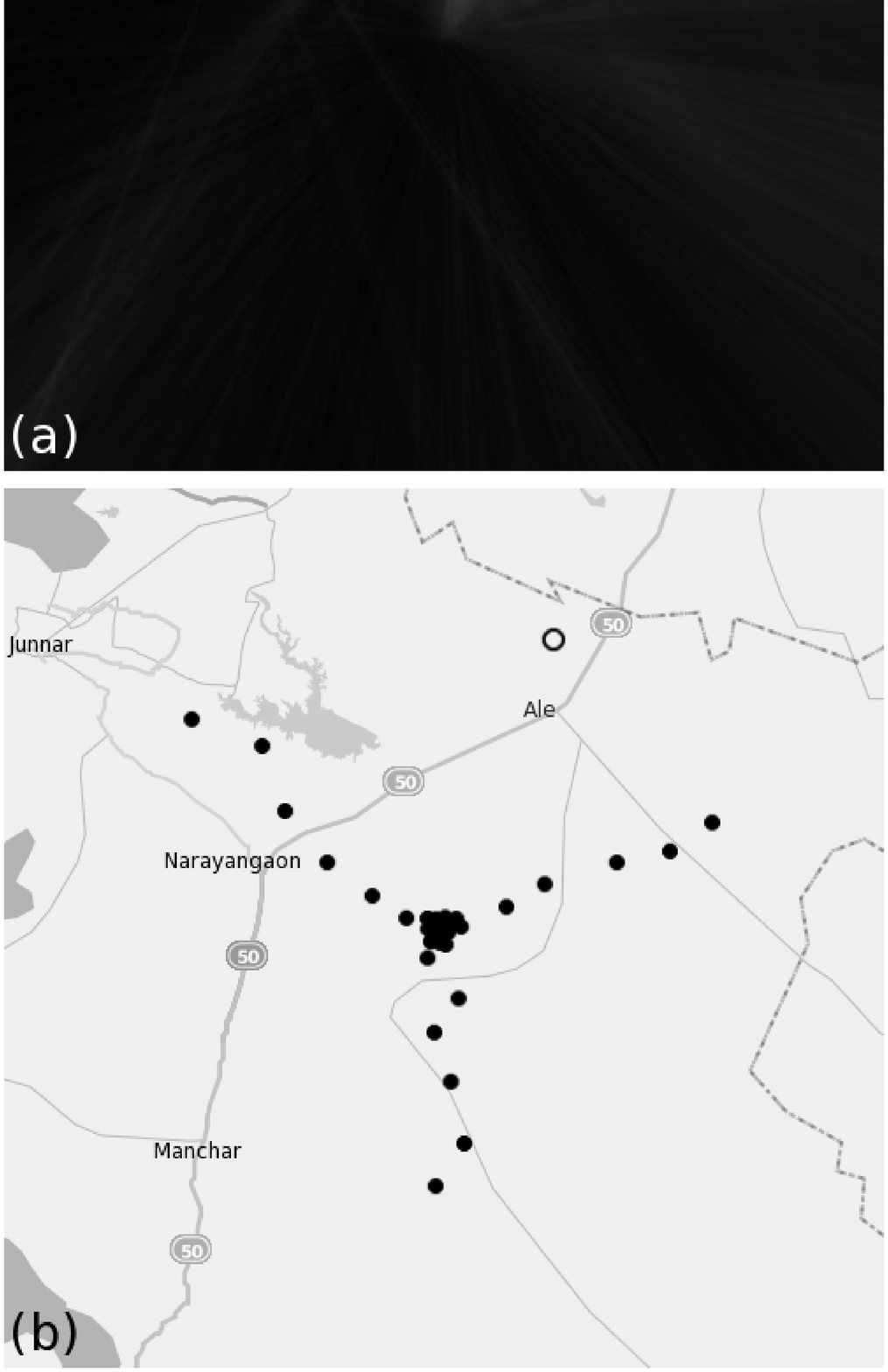}
  \end{center}
  \caption{(a) A near-field image, 
    measuring about 40\,km on each side 
    and centred on GMRT.
    An RFI source identified through large SVD modes 
    is clearly visible as a bright spot in the upper right.
    (b) A map of GMRT covering the same area as in (a). 
    Each black circle indicates an antenna location, 
    and the open circle corresponds to the RFI source, 
    \#14 in Table~\ref{tab:RFI}, 
    identified in (a).}
  \label{fig:RFI-source}
\end{figure}

Candidate sources detected by a single baseline 
appear as a hyperbola of equal light arrival time 
in near-field images of a single SVD mode. 
When a source is detected by many baselines, 
the corresponding hyperbolas intersect at a single point in the image. 
These near-field images, 
an example of which is shown in Fig.~\ref{fig:RFI-source}, 
become the `RFI maps' which are used to isolate interference sources.
For sources with a high enough duty cycle, 
the GPS position given by the RFI maps 
are accurate enough that a handheld yagi antenna 
and portable radio receiver 
can be used to find the precise source. 
With the handheld antenna direction to a source can be determined by ear 
to better than 30 degrees, 
and thereafter localised by successive triangulation.

By comparing the position of a noise transmitter 
relative to the candidate interference source 
both physically and in the RFI map, 
we can confirm whether we have identified the correct source.  
Occasional misidentifications are likely 
due to the intermittent nature of the sources; 
there is no guarantee that the radiating source in the images 
will still be active when an attempt at identification is made. 
Since the beginning of this effort, 
calibration has been improved to locate sources 
within about 100\,m. 
Sources could in principle be located using only the array 
with a precision 10\,m, 
but the accuracy of available maps, GPS equipment, and other factors 
limit the real-world precision.

Candidate RFI sources identified so far include transformers, 
power line junctions, 
and loose wires in contact with power lines. 
Table~\ref{tab:RFI} lists some candidate sources 
recorded on 2008 December 6, and descriptions. 
In February 2009, GMRT began a collaboration effort with MSEDCL, 
which controls the power transmission lines in the area around GMRT, 
to assist us in removing local interference sources. 
A wire hanging over 132\,kV high tension power lines 
(\#1 in Table~\ref{tab:RFI}) 
which was by far the brightest interference source within 20\,km, 
appearing in 30 of the 50 largest eigenmodes, 
was removed in February 2009. 
Other bright sources have also been removed since then. 
Although these sources are still present in the raw data from 2007, 
this procedure has shown that the SVD algorithm applied in the present work
is successful in identifying real RFI sources during analysis.

\begin{table*}
  \caption{List of candidate RFI sources based on observations
    on 2008 December 6, identified by the strongest SVD mode number from
    which the GPS coordinates are derived. The two brightest
    sources, modes number 1 and 9, were removed in early 2009.}
  \begin{tabular}{cllp{12cm}}
    \hline
    Mode & Longitude & Latitude  & Description      \\
    \hline
    1    & 74.034653 & 19.147676 & Conclusively identified as a wire hanging over 132\,kV high tension power lines, clearly visible using the handheld yagi antenna several kilometres away. Wire removed with assistance from MSEDCL on February 26, 2009.\\
    9    & 74.085335 & 19.171797 & Identified with wires hanging from an unused telephone pole, positioned directly under high tension lines.  \\
    13   & 74.034271 & 19.037518 & Source in the area transmitted very intermittently, making identification difficult. Potentially a transformer 20\,m north of coordinate. \\
    14   & 74.096581 & 19.209597 & Identified as a small wire hanging on a 500\,kV power line, removed in February 2010.\\
    15   & 74.120018 & 19.181877 & Two possible sources, both transmission towers, separated by about 300\,m. The one closest to the coordinate is a T-junction of two high voltage lines. \\  
    17   & 74.072952 & 19.088457 & Initially identified to be a small pump approximately 150\,m west of this coordinate, though other candidates include two nearby transformers. \\
    33   & 74.075623 & 19.108976 & Transformer approximately 100\,m northwest, unambiguously radiating at low levels. \\
    \hline
  \end{tabular}
  \label{tab:RFI}
\end{table*}

\subsection{Single-night Images}
\label{sec:single night}

\begin{figure}
  \begin{center}
    \includegraphics[width=\columnwidth]{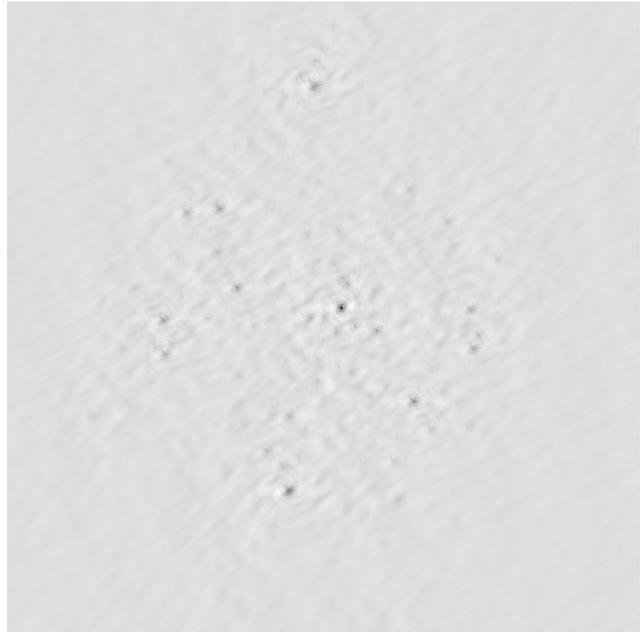}
  \end{center}
  \caption{Dirty sky image made with 8 hours of data from 2007 December 10.
    The maximum $(u,v)$ distance is 2000 
    with grid size of approximately $\Delta u=8$
    giving a field of view of 7.3 degrees on each side. 
    This same $(u,v)$ distance cut is used to calibrate 
    the peak flux of 5C~7.245 to 1.6\,Jy for each day.
    The RMS of this image is 44\,mJy within the primary beam.}
  \label{fig:8hour}
\end{figure}

% Polarization calibration
After RFI removal has been completed in each one hour scan, 
polarization calibration is used to 
combine these into images of an entire night, 
typically of about 8 hours, 
as can be seen in Fig.~\ref{fig:8hour}. 
This step accounts for leakage between 
the left and right polarization signals, 
as well as effects caused by the relative rotation 
of the array and sky source 
by taking into account 
the changing parallactic angle with time. 

Although the array elements very nearly lie in a single plane, 
the rotation of the Earth turns this plane over time 
with respect to the line of sight.  
To correctly stack many hours together, 
one must consider the frequency and directional dependence of the measurement.
One effect to consider is the change of the primary beam with frequency.
For a given frequency, the primary beam is well defined and independent
of $(u,v)$. Deconvolution is difficult, but one can pick the frequency
with the smallest beam and restrict the field of view at other frequencies
to match.
Computationally this can be done by convolving with the ratio of the beams,
which is possible because of the commutativity of the primary beam operators.

However, this is a small effect compared to the effects of a $w$ term.
Typically relative antenna positions are considered to be
on a two dimensional plane with coordinates parametrized by $(u,v)$
in units of the observing wavelength.
Since baselines become non-coplanar over the course of the night,
a third dimension, $w$, must be included,
without which the image becomes blurred.

For a thorough explanation of $w$ term issues,
see for example \citet{Cornwell08} and,
for a more general overview, \citet{Thompson01}.
As the observing frequency changes across the band,
a single baseline samples different points in $(u,v)$ space.
At any given frequency, the $(u,v)$ plane is typically sparsely sampled,
but with many frequency channels
there will often be data at a single $(u,v)$ point for at least a few channels.
One can subtract the average from these,
and under the assumption that foreground sources have the same
spectral index as the calibrator, this removes the foregrounds very well.
Unfortunately, when considering the full $(u,v,w)$ cube,
the data is too sparse for this to work.
Since the $w$ coordinate is dependent on the position in $(u,v)$ space,
and can become very large,
the effects of the $w$ term can not be corrected at different frequencies
in the same way as the beam. 

Strategies for correcting this effect for GMRT are being developed, 
which build on those used in CMB studies \citep{Myers03, Hobson02}, 
but we have not yet applied these to our data. 
In the interim we restrict ourselves to the short baselines 
for which $w$ is small and such corrections are not needed. 
The short baselines are the ones most sensitive to the EoR signal, 
so we retain most of the EoR sensitivity of the array.

The flux scale of the field is set by calibrating
relative to the pulsar in the centre of the field
for every baseline, receiver, and frequency.
Since the pulsar flux is known to be variable,
to set the absolute flux scale we identify the source 
with the highest flux in the sky image, 
and set this peak value equal to the known flux of that source. 
The exact value found depends on the angular resolution, 
determined by the maximum baseline length used. 
For all days analysed in this work, 
the brightest apparent source is the radio galaxy 5C~7.245.
The flux of this source was measured as $215.1\pm12.6$\,mJy at 1415\,MHz
by \citet{Willis76} at the Westerbork Radio Telescope,
and as $685\pm47$\,mJy at 408\,MHz in the 5C catalogue \citep{Pearson78}
for a spectral index of $0.93\pm0.07$.
Following the same argument as \citet{Pen09},
to achieve the best match with other bright sources in the field,
we adopt a value of 1.6\,Jy for 5C~7.245.
Although the flux of this source changes by as much as 10 per cent across the band,
this change is less than that due to the uncertainty in the spectral index.
Additionally, this galaxy has two components separated by 12\,arcsec \citep{Willott01}.
GMRT is capable of resolving this structure with 
$|{\bf b}|\ga2600$, 
where ${\bf b} = (u,v)$ is the distance between the two
antennas in the baseline in units of the observing wavelength.
To calibrate the flux we require a point source 
so that the flux is not spread over multiple image pixels. 
This is achieved by limiting the maximum baseline length used while calibrating 
to $|{\bf b}|\le2000$.
At much smaller $|{\bf b}|$, we become similarly limited by confusion.

Removing the foregrounds adequately is essential
for detecting the EoR signal \citep[see, e.g.,][]{Wang06}
and much work has been done on simulating foregrounds
\citep[e.g.][]{Bowman09, Jelic10}
and designing removal strategies
\citep[][and others]{Morales06, Harker09, Liu09b}.
For a review see \citet{Morales10} and references therein.
To remove foreground sources, 
we take advantage of the fact that the flux of such sources 
do not vary greatly with frequency. 
A signal originating from the EoR will appear as additional
variation on top of the foreground signal.
To model spectrally smooth sources, 
for every timestamp and baseline we subtract a piecewise linear fit 
between the median fluxes in frequency bins of 8, 2, or 0.5\,MHz. 
This is simple to implement and has the advantage over
a polynomial fit of a having straightforward and local window function
which is simple to interpret.
Foregrounds which do not vary over this range are removed,
while features with variability at higher frequencies remain. 
This method results in an upper limit to the EoR signal
since the measured power may still include residual foreground variation.
By subtracting the mean flux over 2\,MHz to remove galactic foregrounds,
noise levels can be lowered to about 2\,mJy
for maps of approximately 10 degrees.
Bright point sources may play a significant role \citep{Datta10, Liu09a, DiMatteo04},
but are not treated separately here.

To first order, this is the same as a boxcar average
\footnote{More precisely it is a boxcar average with a
variable width, and therefore a somewhat different curvature.}
which we can write as
\begin{equation}
\tilde{D}(\nu) = \int D(\nu') w(\nu-\nu') \mathrm{d}\nu'
\end{equation}
\noindent where $D(\nu)$ is the input data,
$\tilde{D}(\nu)$ is the data after filtering,
and the window function is
\begin{equation}
w(x) = \delta(x) - H(x,\Delta\nu) .
\end{equation}
\noindent Here, $\delta(x)$ is the Dirac delta function and $H(x,\Delta\nu)$
is a step function centred at $x$ with width $\Delta\nu$
corresponding to our chosen filter,
and of unit area.

\begin{figure}
  \begin{center}
    \includegraphics[width=\columnwidth]{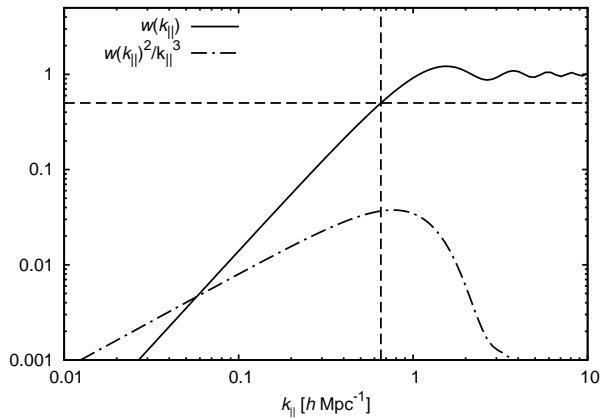}
  \end{center}
  \caption{Fourier transform of our chosen window function
    for $\Delta\nu=0.5$\,MHz (solid line).
    Dashed lines indicate where $w(k_\parallel)=0.5$,
    which sets the minimum $k_{\parallel}$ to which we are sensitive.
    This value scales inversely with $\Delta\nu$.
    The peak of $w(k_\parallel)^2/k_\parallel^3$ (the dot-dashed line)
    indicates the $k_\parallel$ to which we are most sensitive
    under the assumption that the power spectrum is proportional to $k^{-3}$.}
  \label{fig:window}
\end{figure}

In Fourier space, this window function is
\begin{equation}
w(k_\nu) = 1 - \frac{\sin(k_\nu \Delta\nu / 2)}{k_\nu \Delta\nu / 2} ,
\end{equation}
\noindent shown in Fig.~\ref{fig:window},
where we have denoted the wave number as $k_\nu$
to make explicit that it is in units of inverse frequency.
This can be rewritten in terms of $k_\parallel$ using
the fact that
in our redshift range $1\,\mathrm{MHz} \approx 11.6\,h^{-1}\mathrm{Mpc}$.
When converted to units of $h\mathrm{Mpc}^{-1}$, 
we write the wave number as $k_\parallel$
to emphasize that the window function acts on structure 
along the line of sight only.

After filtering in this way, we denote the $k_\nu$ for which $w(k_\nu)=0.5$ 
as corresponding to
the minimum $k_\parallel$ along the line of sight to which we are sensitive,
while smaller $k_\parallel$ are removed by the filter and
will not contribute as strongly to the power spectrum.
The three filters of 8\,MHz, 2\,MHz, and 0.5\,MHz correspond to
minimum $k_\parallel$ of 0.04, 0.16, and 0.65\,$h$Mpc$^{-1}$ respectively.
Since the power spectrum is thought to be proportional to $k^{-3}$ \citep{Iliev08},
the line $w(k_\parallel)^2/k_\parallel^3$ indicates
the $k_\parallel$ to which we are most sensitive.

\begin{figure}
  \begin{center}
    \includegraphics[width=\columnwidth]{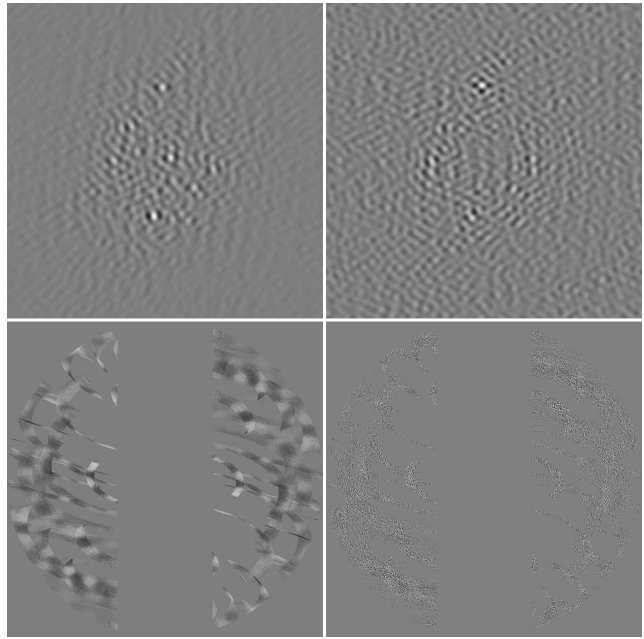}
  \end{center}
  \caption{Data from Dec 10. 
    The top row is the sky image with a 11.4 degree field of view,
    with $|{\bf b}|<200$ and $|u|>60$ 
    binned in the $(u,v)$ plane by $\Delta u=5$.
    The bottom row is the visibilities in the same range 
    with $\Delta u = 0.4$ to show structure.
    The left column is before any foreground subtraction.
    In this image the dominant source is B2\,0825+24 (or 4C\,24.17) 
    just south of the FWHM of the primary beam,
    with a peak value of 2.2\,Jy. 
    RMS within the beam is 343\,mJy.
    The right column is after a 0.5\,MHz subtraction.
    The peak value of this image is 47\,mJy with an RMS of 6.2\,mJy,
    lower by a factor of about 50. If put on the same grey-scale
    as the image without foreground subtraction no features would be visible.
    The dominant source after this filter is 3C\,200,
    well outside the beam.}
  \label{fig:sky200}
\end{figure}

\begin{figure}
  \begin{center}
    \includegraphics[width=\columnwidth]{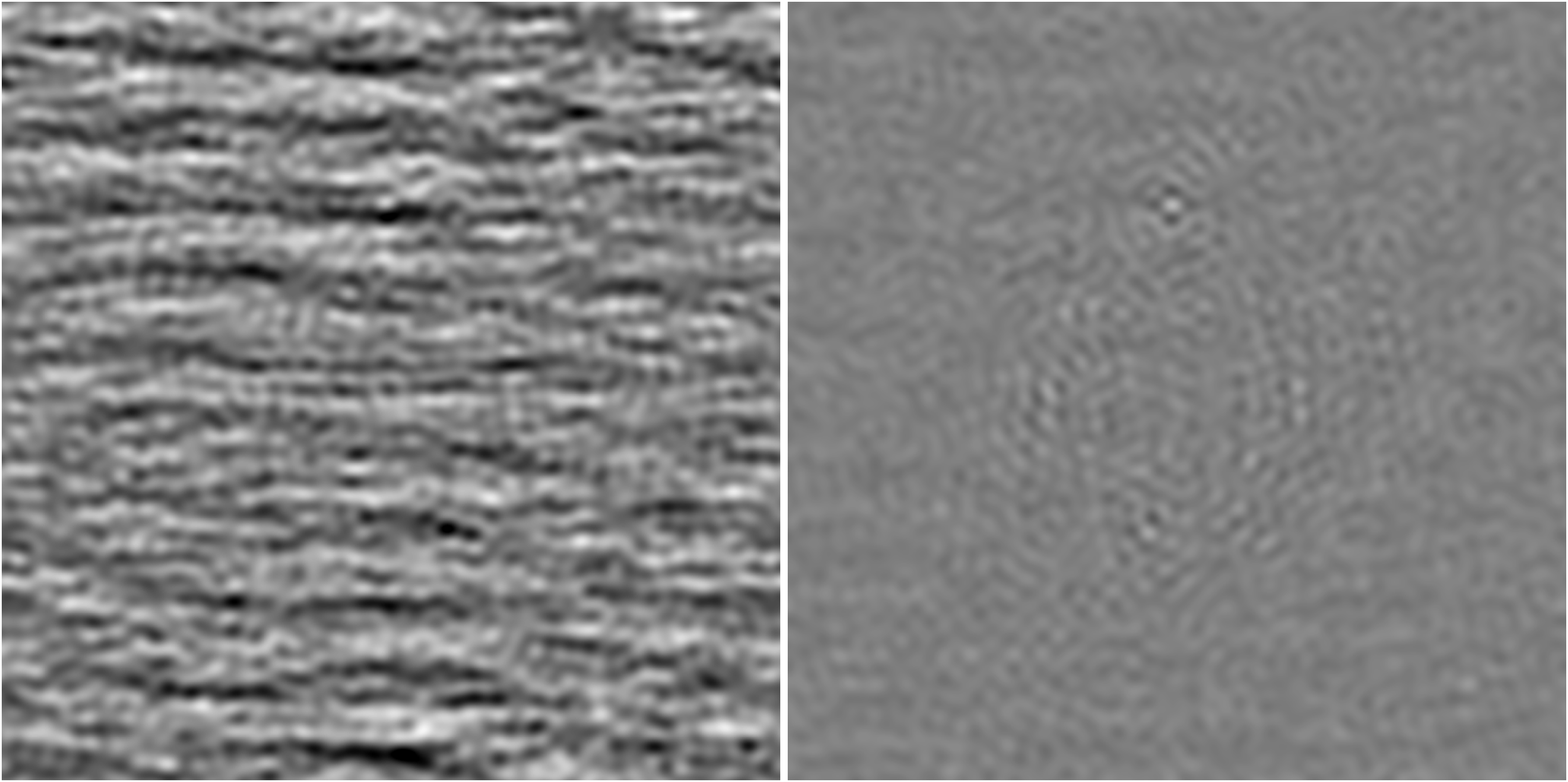}
  \end{center}
  \caption{Dirty sky images before the SVD RFI removal step (left)
    and after the RFI removal (right) with $|{\bf b}|<200$,
    after the 0.5\,MHz foreground subtraction.
    The field of view is 11.4 degrees.
    RFI clearly dominates the map on the left,
    with peak value of 73\,mJy and an RMS of 18\,mJy.
    After RFI removal, the peak drops to 32\,mJy with an RMS of 4.2\,mJy.}
  \label{fig:before-and-after-sky}
\end{figure}

The 0.5\,MHz foreground filter reduces the peak and RMS values 
by a factor of 50, as can be seen in Fig.~\ref{fig:sky200}.
The filter is most effective within the primary beam.
Fig.~\ref{fig:before-and-after-sky} demonstrates the dominance of RFI
in the filtered dirty maps,
and the improvement that the SVD step provides.
Without the SVD, the maps in Fig.~\ref{fig:sky200} would
have been a factor of 4 noisier.

%\begin{figure*}
%\end{figure*}

%%%%%%%%%%%%%%%%%%%%%%%%%%%%%%%%%%%%%%%%%%%%%%%%%%%%%%%%%%%%%%%%%%%%%%%%%%%%%%
\section{Multiday Analysis}
\label{sec:Multiday}
%%%%%%%%%%%%%%%%%%%%%%%%%%%%%%%%%%%%%%%%%%%%%%%%%%%%%%%%%%%%%%%%%%%%%%%%%%%%%%

\subsection{Differences Between Nights}

Since RFI and foreground signals are so much larger than the EoR signal, 
small errors in the subtraction of foregrounds could easily result 
in a spurious signal. To avoid this we use cross-correlations between
multiple days to make a statistical measurement.

To gauge how successful a cross-correlation measurement might be, 
we are interested in the relative similarity of the different nights,
which we can gauge by taking the difference of visibilities.  
Days which subtract well will show mostly noise, 
while days which subtract poorly will still show evidence of bright sources. 
Visibilities are only used when data exists at the same position in both days. 
This serves as a test of the important of ionospheric fluctuations.
Though the ionosphere is calibrated for along the light of sight to the pulsar
in the centre of the field, the ionosphere could change across the field of view.
Pairs of days for which this change away from the field of view is different
will not subtract well.
The results of the subtractions are shown in Table~\ref{tab:diffs}.

It can be seen in Table~\ref{tab:diffs} 
that December 11 gives consistently poor results
in both the unfiltered and filtered images, 
and was thus excluded from all subsequent analysis.  
From these maps we can also conclude that 
RFI is not dominating the differences from day to day. 
If this had been the case, 
since RFI is typically isolated in $(u,v)$ space, 
it would be visible across the whole sky image, 
including far outside the primary beam. 
However, it can be seen in Fig.~\ref{fig:sky200} that
structure decreases rapidly away from the centre of the field, 
indicating an astronomical source. 
This is true of all subtraction pairs.

\begin{table}
    \caption{Peak flux and RMS of the difference of each two day pair
      available, in mJy, with a maximum $(u,v)$ distance of 600, which
      corresponds to a maximum baseline of 1.2\,km.  All
      values are in mJy. To remove foregrounds, a 2\,MHz linear filter
      was applied.}
    \begin{tabular}{cc|rr|rr}
      \hline
           &   & \multicolumn{2}{|c|}{Unfiltered}  &  \multicolumn{2}{c}{With 2\,MHz filter} \\
      \hline
         \multicolumn{2}{c|}{Subtracted pair}  & Peak Flux  &  RMS & Peak Flux &  RMS  \\ 
      \hline
  Dec 10  &  Dec 11  &  1856.1  &   227.3  &   137.9  &    19.9 \\
  Dec 10  &  Dec 14  &   888.4  &   185.6  &    62.5  &    11.6 \\
  Dec 10  &  Dec 16  &   278.4  &    49.8  &    27.1  &     3.2 \\
  Dec 10  &  Dec 17  &   447.9  &    64.9  &    26.2  &     3.6 \\
  Dec 10  &  Dec 18  &   548.3  &    70.4  &    32.5  &     4.2 \\
  Dec 11  &  Dec 14  &  1037.8  &   145.1  &   256.1  &    56.5 \\
  Dec 11  &  Dec 16  &  2148.3  &   256.7  &   158.1  &    23.9 \\
  Dec 11  &  Dec 17  &  2528.2  &   318.5  &   106.1  &    21.1 \\
  Dec 11  &  Dec 18  &  2997.5  &   356.1  &   159.1  &    12.4 \\
  Dec 14  &  Dec 16  &  1221.5  &   193.9  &    50.2  &     7.0 \\
  Dec 14  &  Dec 17  &  1311.1  &   257.2  &    50.4  &     7.6 \\
  Dec 14  &  Dec 18  &  1433.5  &   233.0  &    67.1  &     4.9 \\
  Dec 16  &  Dec 17  &   202.0  &    78.9  &    23.8  &     2.8 \\
  Dec 16  &  Dec 18  &   224.2  &    31.2  &    24.4  &     3.4 \\
  Dec 17  &  Dec 18  &   206.9  &    60.6  &    24.6  &     3.2 \\
      \hline
    \end{tabular}
    \label{tab:diffs}
\end{table}

\subsection{Cross-correlations and Power Spectra}
\label{sec:cc-and-ps}

\begin{figure}
  \begin{center}
    \includegraphics[width=\columnwidth]{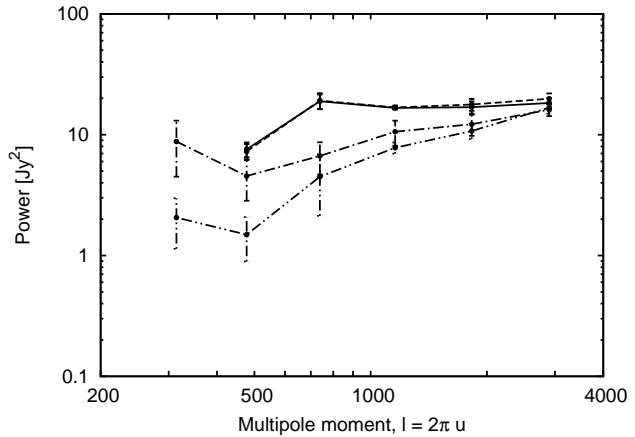}
  \end{center}
  \caption{Comparison of the cross-power of December 10 with all other days
    under four different conditions.
    The dot-dashed and double-dot-dashed lines are 
    before and after the SVD RFI removal, 
    respectively, including all $u$.
    It can be seen that power is lost in the SVD.
    Similarly, the solid and dashed lines are before and after the SVD step, respectively,
    this time with a $|u|>60$ limit imposed.
    In this case the two lines are almost identical,
    meaning the SVD had little effect on the total power.}
  \label{fig:umin60}
\end{figure}

The power spectrum of sky structure can be determined directly from the visibilities \citep{Zaldarriaga04}.  
To find the cross-power of two days, 
we take the product of the visibilities after gridding with a cell spacing
equal to the size of the beam ($\Delta u = 20$).
Calculating the power spectrum from the visibilities
instead of from the correlations in the sky image
takes advantage of the fact that the visibilities have a nearly
diagonal correlation matrix in the noise \citep{White99}.
We then find the weighted average of visibilities 
in annuli of $(u,v)$ space
which gives the power in units of Jy$^2$.
Since the amount of data at large $|{\bf b}|$ decreases,
we increase the width of each successive annuli by 60 per cent with increasing $|{\bf b}|$.
The smallest bin width is equal to the size of the beam.
This prevents large artificial variability in power 
due to sparse sampling. 
Visibilities are weighted `naturally', by the inverse of the noise. 
Since we expect our sensitivity to the EoR signal 
to diminish rapidly with increasing baseline length, 
we look only at the first few points, 
averaged over all possible cross-correlations. 
Additionally, 
it is known that the SVD will introduce a loss of power at low $|u|$. 
To avoid this, we impose a limit of $|u|>60$ when taking the cross-correlations, 
determined by requiring that the power spectrum before and after the SVD 
differ by less than 1$\sigma$
as shown in Fig.~\ref{fig:umin60}.
The part of the $(u,v)$ plane that is lost with this cut
can be seen in Fig.~\ref{fig:before-and-after-vis}.

\begin{figure}
  \begin{center}
    \includegraphics[width=\columnwidth]{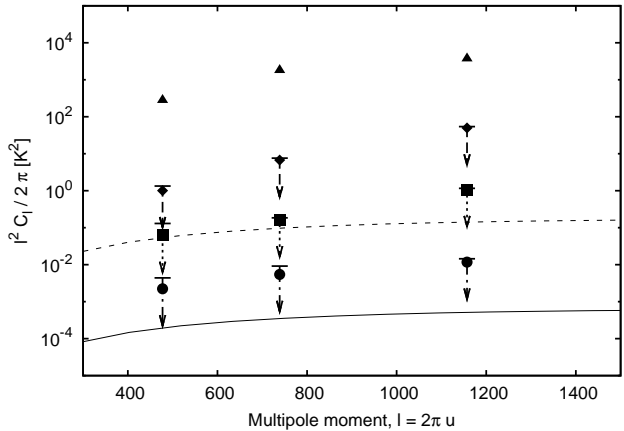}
  \end{center}
  \caption{Average power spectrum in units of K$^2$ of all
    combinations of days, excluding December 11, as a function of
    the multipole moment $l$. Each point is shown with a $2\sigma$
    upper limit derived from a bootstrap error analysis,
    which is in most cases smaller than the size of the point.
    The points are logarithmically spaced as described in the text,
    from left to right covering the ranges 
    %$60<|{\bf b}|<92$, $92<|{\bf b}|<143$, and $143<|{\bf b}|<225$.
    $377<l<578$, $578<l<899$, and $899<l<1414$.
    Triangles are the power before subtracting foregrounds,
    diamonds are after 8\,MHz mean subtraction,
    squares are after 2\,MHz mean subtraction,
    and circles are after 0.5\,MHz subtraction.
    The curved solid line is the theoretical EoR signal from
    \citet{Jelic08}, and the dashed line is the theoretical EoR signal
    with a cold absorbing IGM as described in the text.}
  \label{fig:power-K}
\end{figure}

The power spectrum of the cross-correlation can
be converted to units of K$^2$ using
\[
\frac{l ^2}{2 \pi} C_{l} \left| _{l = 2\pi |{\bf b}|} \right. =
  \left( \frac{|{\bf b}|}{11.9} \frac{3.3^\circ}{\theta_b} \right)^2
  \left( \frac{150\,\mathrm{MHz}}{\nu} \right)^4
  \left< \left| \frac{V({\bf b})}{\mathrm{Jy}} \right| ^2 \right> \mathrm{K}^2
\]
\noindent where $l=2 \pi |{\bf b}|$,
${\bf b} = (u,v)$ is the visibility coordinate in units of wavelength,
$C_{l}$ is the power measured in K$^2$,
$\theta_b$ is the primary beam size,
and $\nu$ is the wavelength \citep{Pen09}.
The quantity in the angle-brackets on the right
is equal to the power in Jy$^2$ found above.
This conversion is written in terms of the GMRT observations
with a primary beam of $\theta_b = 3.3^\circ$ and $\nu=150$\,MHz.
If gridding in the $(u,v)$ plane is too fine, the data become noisy,
while very coarse gridding requires the assumption that
the data are constant across the whole cell.
As mentioned, we use a gridding equal to the size of the beam.
A fully optimized estimate would require
a maximum likelihood code
which is being worked on in the context of the Cosmic Background Imager (CBI) gridder \citep{Myers03}.
Fig.~\ref{fig:power-K} shows the weighted-average of all cross-correlation pairs, 
excluding December 11.
\citet{Bernardi09} reported a power spectrum of foregrounds without subtraction
in the galactic plane at a level comparable to our measurement
both here and in \citet{Pen09}.

We have plotted the power spectra with $2\sigma$ bootstrap errors \citep{Efron79}
which were derived as follows:
Using five days of data, there are ten possible cross-correlations.
From these, ten are randomly sampled, with replacement,
resulting in a slightly different power spectrum. 
This is repeated $10^4$ times, and the variance on
this set of power spectra is calculated to give the
error on the original. 
Formally this quantifies the error when taking
independent samples of a statistical distribution.
In our case this is not a rigorous error,
but a suitable straightforward estimate
given the main complications discussed earlier.

\subsection{Comparison to Models}
\label{sec:comparison}

Fig.~\ref{fig:power-K} can be compared 
to simulated results from the Low Frequency Array (LOFAR) EoR
project in \citet{Jelic08}, which assumes $T_\mathrm{s} \gg
T_\mathrm{CMB}$. At low $l$, their simulated EoR signal is
approximately (10\,mK)$^2$, while our lowest point
with a similar 0.5\,MHz bandwidth filter is
(50\,mK)$^2$ with a $2\sigma$ upper limit of (70\,mK)$^2$.
These results are comparable to
the sensitivities LOFAR expects after 400\,hours of the EoR
project. We have also considered the case where reheating of the IGM
does not occur, so the spin temperature remains coupled to the kinetic
temperature of the gas \citep{Ciardi03}.
In this case, the IGM cools adiabatically
after decoupling from the CMB at $z \approx 150$. The temperature
fluctuations scale with $(1+T_\mathrm{CMB}/T_\mathrm{s})$, and the
power scales with the same factor squared. Using
$T_\mathrm{k}=T_\mathrm{CMB}(1+z)/150$ at $z=8.6$, the power becomes
approximately 275 times larger.
This line is shown in Fig.~\ref{fig:power-K},
and is comparable to the data.
Scaling up the warm IGM power spectrum from \citet{Iliev08}
or \citet{Jelic08}
in this way is a reasonable approximation to what the
expected signal in a cold IGM.
For a more detailed studies of the signal in such an absorption regime,
we point the reader to \citet{Baek09} and \citet{Baek10}.

\begin{figure}
  \begin{center}
    \includegraphics[width=\columnwidth]{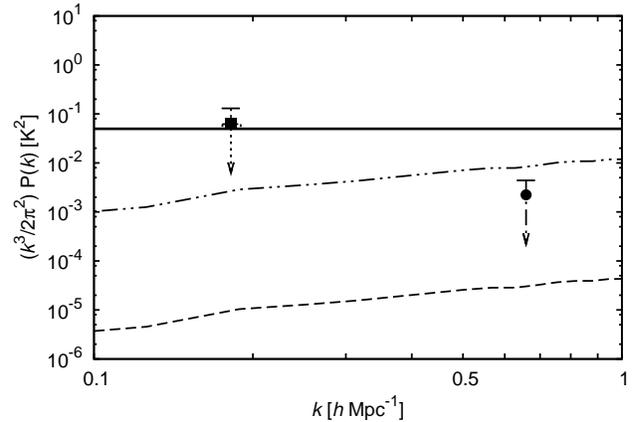}
    \end{center}
  \caption{3D power spectrum for the same data shown in Fig.~\ref{fig:power-K}, using
    $k^2=k_\parallel^2+k_\perp^2$.
    This is dominated by $k_\parallel$, so the bin
    width in $k_\perp$ does not influence the horizontal
    position of the limits.
    The strongest constraints from the 2\,MHz and
    0.5\,MHz filters are shown (square and circle, respectively).
    Upper limits are $2\sigma$ bootstrap errors.
    Three possible signals are shown.
    The dashed line is the prediction from \citet{Iliev08}
    and the double-dot-dashed line is the same for a cold IGM.
    The solid line comes from the single-scale bubble model
    as described in the text for a cold IGM,
    using $k=2.5/R$ to show the maximum power at all $k$.
    For the two points shown, the bubble diametres which achieve
    this maximum power are 27 and 7.4\,$h^{-1}$Mpc respectively.
    Only the 0.5\,MHz point imposes a limit on the diametre.
    For a warm IGM case, this signal would be reduced by the same factor
    as in the two dashed lines.}
  \label{fig:power-3D}
\end{figure}

The power spectrum of reionization is intrinsically three dimensional \citep{Morales04}.
The strongest constraints on the
3D power spectrum for the $\Delta\nu=2$ and $0.5$\,MHz foreground filter case
are shown in Fig.~\ref{fig:power-3D}.
This uses $k^2=k_\parallel^2+k_\perp^2$,
where $k_\parallel$ is given by the windowing function of the filter
and $k_\perp = l / 6\,h^{-1} \mathrm{Gpc}$.  When comparing this to the prediction
from \citet{Iliev08}, one should note that our windowing function
will also reduce the predicted signal by at most a factor of two.

We also consider an idealized case in which the ionized bubbles
during reionization are of uniform scale and non-overlapping.
Then for a given $k$ there will be a characteristic bubble radius $R$
at which the power is maximized.
By taking the 3D Fourier transform of a perfectly ionized bubble,
and requiring that the universe is 50 per cent ionized,
one can show the power is
given by
\begin{equation}
  \label{eqn:single-scale-power}
  \frac{k^3}{2\pi^2} P(k) = \frac{3T_\mathrm{b}^2}{2\pi}(kR)^3 \left[ \frac{\cos kR}{(kR)^2} - \frac{\sin kR}{(kR)^3} \right]^2 
\end{equation}
and is maximized when $kR \approx 2.5$.
In this case, $k^3/(2\pi^2) P(k) \approx T_\mathrm{b}^2/5$,
where $T_\mathrm{b}$ is the brightness temperature
$\approx 30$\,mK in an X-ray heated IGM
or almost -500\,mK in a cold absorbing IGM.
This signal would be more than an order of magnitude larger
than the predictions by e.g. \citet{Iliev08} or \citet{Jelic08}.
In Fig.~\ref{fig:power-3D} we have included the
power spectrum from this model with $k$ chosen to maximize the power
in the range of interest. The data currently impose a limit
on the size of bubbles in this single-scale model.
Our upper limits
with a 0.5\,MHz foreground subtraction
rule out bubbles with diametres from 2.2 to 12.4\,$h^{-1}$Mpc
in the redshift range $8.1<z<9.2$.
The cold IGM constraint is applicable even in the case of simulations,
since only UV photons are included
which themselves do not heat the IGM.

%%%%%%%%%%%%%%%%%%%%%%%%%%%%%%%%%%%%%%%%%%%%%%%%%%%%%%%%%%%%%%%%%%%%%%%%%%%
\section{Conclusion}
\label{sec:Conclusion}
%%%%%%%%%%%%%%%%%%%%%%%%%%%%%%%%%%%%%%%%%%%%%%%%%%%%%%%%%%%%%%%%%%%%%%%%%%%

The data analysis has been completed on six days from December 2007
with a noise level of approximately 2\,mJy on most nights.  
The SVD removal strategies for broadband RFI used lower noise
by a factor of 4 in temperature, or 16 in power,
which flagging alone cannot achieve.
We have also tested for ionospheric variations and found
that our pulsar calibration is sufficient for dealing with these effects.
After RFI removal and foreground subtraction, we have measured a
power spectrum which represents a new upper limit on the 21\,cm
brightness temperature fluctuations during the epoch of reionization.
These results can be used to constrain assumptions about the state of the 
IGM at these times, particularly in the case of a Ly-$\alpha$
pumped, but cold, IGM.  

The previous best limit on 21\,cm signal at comparable redshift was by
\citet{Bebbington86}, who reported no features down to 5\,K at $z=8.4$. 
\citet{Parsons10} have reported a similar limit of about 5\,K using PAPER.
The upper limit we present here
is approximately 70\,mK on the variance in 21\,cm brightness temperature at $z=8.6$,
almost two orders of magnitude better than these previous limits.
Residual foreground contamination and RFI may still be contributing to this power,
but the EoR signal can not be larger than this.

The analysis of additional observations of the B0823+26 field made
since 2007, containing approximately 2 times more data than has been
treated in this work, is ongoing. We expect to continue to improve these results, and the
planned addition of a second field will strengthen this statistical
measurement. The dominant uncertainty source remains RFI and foreground modelling.

These results provide a first-look at the progress made at GMRT in detecting EoR. 
The GMRT-EoR team has been continuing observations,
most recently with an additional 150 hours allocated in the summer of 2010.
We continue to improve both the system temperature of GMRT
antennas and the RFI monitoring and removal strategies and expect 
to improve upon these results as the analysis of the newer data progresses.

%%%%%%%%%%%%%%%%%%%%%%%%%%%%%%%%%%%%%%%%%%%%%%%%%%%%%%%%%%%%%%%%%%%%%%%%%%%
\section{Acknowledgments}
\label{sec:Acknowledgments}
%%%%%%%%%%%%%%%%%%%%%%%%%%%%%%%%%%%%%%%%%%%%%%%%%%%%%%%%%%%%%%%%%%%%%%%%%%%

We thank Chris Hirata for his contributions to the analysis pipeline 
and a reading of the paper, and the anonymous reviewer for many
useful comments.
We also thank the staff of the GMRT that made these observations
possible. GMRT is run by the National Centre for Radio Astrophysics of
the Tata Institute of Fundamental Research.
The computations were performed on CITA's Sunnyvale clusters
which are funded by the Canada Foundation for Innovation,
the Ontario Innovation Trust,
and the Ontario Research Fund.
The work of UP is supported by the
National Science and Engineering Research Council of Canada.
The map image in Fig.~\ref{fig:RFI-source}
is copyright by OpenStreetMap\footnote{\url{http://www.openstreetmap.org/}} contributors
and used under license\footnote{\url{http://creativecommons.org/licenses/by-sa/2.0/}}.

\bibliography{Upper_Limit_v3}

\begin{thebibliography}{}

\bibitem[\protect\citeauthoryear{{Alvarez}, {Pen} \& {Chang}}{{Alvarez}
  et~al.}{2010}]{Alvarez10}
{Alvarez} M.~A.,  {Pen} U.,    {Chang} T.,  2010, ApJ, 723, L17

\bibitem[\protect\citeauthoryear{{Ananthakrishnan}}{{Ananthakrishnan}}{1995}]{%
GMRT}
{Ananthakrishnan} S.,  1995, Journal of Astrophysics and Astronomy Supplement,
  16, 427

\bibitem[\protect\citeauthoryear{{Baek}, {Di Matteo}, {Semelin}, {Combes} \&
  {Revaz}}{{Baek} et~al.}{2009}]{Baek09}
{Baek} S.,  {Di Matteo} P.,  {Semelin} B.,  {Combes} F.,    {Revaz} Y.,  2009,
  A\&A, 495, 389

\bibitem[\protect\citeauthoryear{{Baek}, {Semelin}, {Di Matteo}, {Revaz} \&
  {Combes}}{{Baek} et~al.}{2010}]{Baek10}
{Baek} S.,  {Semelin} B.,  {Di Matteo} P.,  {Revaz} Y.,    {Combes} F.,  2010,
  preprint (astro-ph/1003.0834)

\bibitem[\protect\citeauthoryear{{Bebbington}}{{Bebbington}}{1986}]{Bebbington%
86}
{Bebbington} D.~H.~O.,  1986, MNRAS, 218, 577

\bibitem[\protect\citeauthoryear{{Becker}, {Fan}, {White}, {Strauss},
  {Narayanan}, {Lupton}, {Gunn}, {Annis} \& {et al}}{{Becker}
  et~al.}{2001}]{Becker01}
{Becker} R.~H. {et al.}, 2001, AJ, 122,
  2850

\bibitem[\protect\citeauthoryear{{Bernardi}, {de Bruyn}, {Brentjens}, {Ciardi},
  {Harker}, {Jeli{\'c}}, {Koopmans}, {Labropoulos}, {Offringa}, {Pandey},
  {Schaye}, {Thomas}, {Yatawatta} \& {Zaroubi}}{{Bernardi}
  et~al.}{2009}]{Bernardi09}
{Bernardi} G.,  {de Bruyn} A.~G.,  {Brentjens} M.~A.,  {Ciardi} B.,  {Harker}
  G.,  {Jeli{\'c}} V.,  {Koopmans} L.~V.~E.,  {Labropoulos} P.,  {Offringa} A.,
   {Pandey} V.~N.,  {Schaye} J.,  {Thomas} R.~M.,  {Yatawatta} S.,    {Zaroubi}
  S.,  2009, A\&A, 500, 965

\bibitem[\protect\citeauthoryear{{Bowman}, {Morales} \& {Hewitt}}{{Bowman}
  et~al.}{2006}]{Bowman06}
{Bowman} J.~D.,  {Morales} M.~F.,    {Hewitt} J.~N.,  2006, ApJ, 638, 20

\bibitem[\protect\citeauthoryear{{Bowman}, {Morales} \& {Hewitt}}{{Bowman}
  et~al.}{2009}]{Bowman09}
{Bowman} J.~D.,  {Morales} M.~F.,    {Hewitt} J.~N.,  2009, ApJ, 695, 183

\bibitem[\protect\citeauthoryear{{Chen} \& {Miralda-Escud{\'e}}}{{Chen} \&
  {Miralda-Escud{\'e}}}{2004}]{Chen04}
{Chen} X.,  {Miralda-Escud{\'e}} J.,  2004, ApJ, 602, 1

\bibitem[\protect\citeauthoryear{{Chuzhoy} \& {Shapiro}}{{Chuzhoy} \&
  {Shapiro}}{2006}]{Chuzhoy06}
{Chuzhoy} L.,  {Shapiro} P.~R.,  2006, ApJ, 651, 1

\bibitem[\protect\citeauthoryear{{Ciardi} \& {Madau}}{{Ciardi} \&
  {Madau}}{2003}]{Ciardi03}
{Ciardi} B.,  {Madau} P.,  2003, ApJ, 596, 1

\bibitem[\protect\citeauthoryear{{Cooray}, {Li} \& {Melchiorri}}{{Cooray}
  et~al.}{2008}]{Cooray08}
{Cooray} A.,  {Li} C.,    {Melchiorri} A.,  2008, Phys.~Rev.~D, 77, 103506

\bibitem[\protect\citeauthoryear{{Cornwell}, {Golap} \& {Bhatnagar}}{{Cornwell}
  et~al.}{2008}]{Cornwell08}
{Cornwell} T.~J.,  {Golap} K.,    {Bhatnagar} S.,  2008, IEEE Journal of
  Selected Topics in Signal Processing, Vol.~2, Issue 5, p.647-657, 2, 647

\bibitem[\protect\citeauthoryear{{Dalal}, {Pen} \& {Seljak}}{{Dalal}
  et~al.}{2010}]{Dalal10}
{Dalal} N.,  {Pen} U.,    {Seljak} U.,  2010, preprint (astro-ph/1009.4704)

\bibitem[\protect\citeauthoryear{{Datta}, {Bowman} \& {Carilli}}{{Datta}
  et~al.}{2010}]{Datta10}
{Datta} A.,  {Bowman} J.~D.,    {Carilli} C.~L.,  2010, ApJ, 724, 526

\bibitem[\protect\citeauthoryear{{Di Matteo}, {Ciardi} \& {Miniati}}{{Di
  Matteo} et~al.}{2004}]{DiMatteo04}
{Di Matteo} T.,  {Ciardi} B.,    {Miniati} F.,  2004, MNRAS, 355, 1053

\bibitem[\protect\citeauthoryear{{Dijkstra}, {Haiman} \& {Loeb}}{{Dijkstra}
  et~al.}{2004}]{Dijkstra04}
{Dijkstra} M.,  {Haiman} Z.,    {Loeb} A.,  2004, ApJ, 613, 646

\bibitem[\protect\citeauthoryear{{Djorgovski}, {Castro}, {Stern} \&
  {Mahabal}}{{Djorgovski} et~al.}{2001}]{Djorgovski01}
{Djorgovski} S.~G.,  {Castro} S.,  {Stern} D.,    {Mahabal} A.~A.,  2001, ApJ,
  560, L5

\bibitem[\protect\citeauthoryear{{Dunkley}, {Komatsu}, {Nolta}, {Spergel},
  {Larson}, {Hinshaw}, {Page}, {Bennett}, {Gold}, {Jarosik}, {Weiland},
  {Halpern}, {Hill}, {Kogut}, {Limon}, {Meyer}, {Tucker}, {Wollack} \&
  {Wright}}{{Dunkley} et~al.}{2009}]{Dunkley09}
{Dunkley} J.,  {Komatsu} E.,  {Nolta} M.~R.,  {Spergel} D.~N.,  {Larson} D.,
  {Hinshaw} G.,  {Page} L.,  {Bennett} C.~L.,  {Gold} B.,  {Jarosik} N.,
  {Weiland} J.~L.,  {Halpern} M.,  {Hill} R.~S.,  {Kogut} A.,  {Limon} M.,
  {Meyer} S.~S.,  {Tucker} G.~S.,  {Wollack} E.,    {Wright} E.~L.,  2009,
  ApJS, 180, 306

\bibitem[\protect\citeauthoryear{{Efron}}{{Efron}}{1979}]{Efron79}
{Efron} B.,  1979, Ann. Statist., 7, 1

\bibitem[\protect\citeauthoryear{{Fan}, {Narayanan}, {Strauss}, {White},
  {Becker}, {Pentericci} \& {Rix}}{{Fan} et~al.}{2002}]{Fan02}
{Fan} X.,  {Narayanan} V.~K.,  {Strauss} M.~A.,  {White} R.~L.,  {Becker}
  R.~H.,  {Pentericci} L.,    {Rix} H.,  2002, AJ, 123, 1247

\bibitem[\protect\citeauthoryear{{Fan}, {Strauss}, {Becker}, {White}, {Gunn},
  {Knapp}, {Richards}, {Schneider}, {Brinkmann} \& {Fukugita}}{{Fan}
  et~al.}{2006}]{Fan06}
{Fan} X.,  {Strauss} M.~A.,  {Becker} R.~H.,  {White} R.~L.,  {Gunn} J.~E.,
  {Knapp} G.~R.,  {Richards} G.~T.,  {Schneider} D.~P.,  {Brinkmann} J.,
  {Fukugita} M.,  2006, AJ, 132, 117

\bibitem[\protect\citeauthoryear{{Field}}{{Field}}{1959}]{Field59}
{Field} G.~B.,  1959, ApJ, 129, 536

\bibitem[\protect\citeauthoryear{{Furlanetto}, {Lidz}, {Loeb}, {McQuinn},
  {Pritchard}, {Shapiro}, {Alvarez} \& {et al}}{{Furlanetto}
  et~al.}{2009}]{Furlanetto09}
{Furlanetto} S.~R. {et al}, 2009, in astro2010: The
  Astronomy and Astrophysics Decadal Survey Vol.~2010 of Astronomy, {Cosmology
  from the Highly-Redshifted 21 cm Line}.
pp 82--+

\bibitem[\protect\citeauthoryear{{Furlanetto}, {Oh} \& {Briggs}}{{Furlanetto}
  et~al.}{2006}]{Furlanetto06}
{Furlanetto} S.~R.,  {Oh} S.~P.,    {Briggs} F.~H.,  2006, Phys.~Rep., 433, 181

\bibitem[\protect\citeauthoryear{{Furlanetto}, {Zaldarriaga} \&
  {Hernquist}}{{Furlanetto} et~al.}{2004}]{Furlanetto04}
{Furlanetto} S.~R.,  {Zaldarriaga} M.,    {Hernquist} L.,  2004, ApJ, 613, 1

\bibitem[\protect\citeauthoryear{{Gunn} \& {Peterson}}{{Gunn} \&
  {Peterson}}{1965}]{Gunn65}
{Gunn} J.~E.,  {Peterson} B.~A.,  1965, ApJ, 142, 1633

\bibitem[\protect\citeauthoryear{{Guo}, {Wu}, {Xu} \& {Gu}}{{Guo}
  et~al.}{2009}]{Guo09}
{Guo} Q.,  {Wu} X.,  {Xu} H.~G.,    {Gu} J.~H.,  2009, ApJ, 693, 1000

\bibitem[\protect\citeauthoryear{{Harker}, {Zaroubi}, {Bernardi}, {Brentjens},
  {de Bruyn}, {Ciardi}, {Jeli{\'c}}, {Koopmans}, {Labropoulos}, {Mellema},
  {Offringa}, {Pandey}, {Pawlik}, {Schaye}, {Thomas} \& {Yatawatta}}{{Harker}
  et~al.}{2010}]{Harker10}
{Harker} G.,  {Zaroubi} S.,  {Bernardi} G.,  {Brentjens} M.~A.,  {de Bruyn}
  A.~G.,  {Ciardi} B.,  {Jeli{\'c}} V.,  {Koopmans} L.~V.~E.,  {Labropoulos}
  P.,  {Mellema} G.,  {Offringa} A.,  {Pandey} V.~N.,  {Pawlik} A.~H.,
  {Schaye} J.,  {Thomas} R.~M.,    {Yatawatta} S.,  2010, MNRAS, 405, 2492

\bibitem[\protect\citeauthoryear{{Harker}, {Zaroubi}, {Bernardi}, {Brentjens},
  {de Bruyn}, {Ciardi}, {Jeli{\'c}}, {Koopmans}, {Labropoulos}, {Mellema},
  {Offringa}, {Pandey}, {Schaye}, {Thomas} \& {Yatawatta}}{{Harker}
  et~al.}{2009}]{Harker09}
{Harker} G.,  {Zaroubi} S.,  {Bernardi} G.,  {Brentjens} M.~A.,  {de Bruyn}
  A.~G.,  {Ciardi} B.,  {Jeli{\'c}} V.,  {Koopmans} L.~V.~E.,  {Labropoulos}
  P.,  {Mellema} G.,  {Offringa} A.,  {Pandey} V.~N.,  {Schaye} J.,  {Thomas}
  R.~M.,    {Yatawatta} S.,  2009, MNRAS, 397, 1138

\bibitem[\protect\citeauthoryear{{Hobbs}, {Lyne}, {Kramer}, {Martin} \&
  {Jordan}}{{Hobbs} et~al.}{2004}]{Hobbs04}
{Hobbs} G.,  {Lyne} A.~G.,  {Kramer} M.,  {Martin} C.~E.,    {Jordan} C.,
  2004, MNRAS, 353, 1311

\bibitem[\protect\citeauthoryear{{Hobson} \& {Maisinger}}{{Hobson} \&
  {Maisinger}}{2002}]{Hobson02}
{Hobson} M.~P.,  {Maisinger} K.,  2002, MNRAS, 334, 569

\bibitem[\protect\citeauthoryear{{Hogan} \& {Rees}}{{Hogan} \&
  {Rees}}{1979}]{Hogan79}
{Hogan} C.~J.,  {Rees} M.~J.,  1979, MNRAS, 188, 791

\bibitem[\protect\citeauthoryear{{Iliev}, {Mellema}, {Pen}, {Merz}, {Shapiro}
  \& {Alvarez}}{{Iliev} et~al.}{2006}]{Iliev06}
{Iliev} I.~T.,  {Mellema} G.,  {Pen} U.,  {Merz} H.,  {Shapiro} P.~R.,
  {Alvarez} M.~A.,  2006, MNRAS, 369, 1625

\bibitem[\protect\citeauthoryear{{Iliev}, {Mellema}, {Pen}, {Bond} \&
  {Shapiro}}{{Iliev} et~al.}{2008}]{Iliev08}
{Iliev} I.~T.,  {Mellema} G.,  {Pen} U.-L.,  {Bond} J.~R.,    {Shapiro} P.~R.,
  2008, MNRAS, 384, 863

\bibitem[\protect\citeauthoryear{{Jeli{\'c}}, {Zaroubi}, {Labropoulos},
  {Bernardi}, {de Bruyn} \& {Koopmans}}{{Jeli{\'c}} et~al.}{2010}]{Jelic10}
{Jeli{\'c}} V.,  {Zaroubi} S.,  {Labropoulos} P.,  {Bernardi} G.,  {de Bruyn}
  A.~G.,    {Koopmans} L.~V.~E.,  2010, MNRAS, pp 1369--+

\bibitem[\protect\citeauthoryear{{Jeli{\'c}}, {Zaroubi}, {Labropoulos},
  {Thomas}, {Bernardi}, {Brentjens}, {de Bruyn}, {Ciardi}, {Harker},
  {Koopmans}, {Pandey}, {Schaye} \& {Yatawatta}}{{Jeli{\'c}}
  et~al.}{2008}]{Jelic08}
{Jeli{\'c}} V.,  {Zaroubi} S.,  {Labropoulos} P.,  {Thomas} R.~M.,  {Bernardi}
  G.,  {Brentjens} M.~A.,  {de Bruyn} A.~G.,  {Ciardi} B.,  {Harker} G.,
  {Koopmans} L.~V.~E.,  {Pandey} V.~N.,  {Schaye} J.,    {Yatawatta} S.,  2008,
  MNRAS, 389, 1319

\bibitem[\protect\citeauthoryear{{Kassim}, {Lazio}, {Ray}, {Crane}, {Hicks},
  {Stewart}, {Cohen} \& {Lane}}{{Kassim} et~al.}{2004}]{Kassim04}
{Kassim} N.~E.,  {Lazio} T.~J.~W.,  {Ray} P.~S.,  {Crane} P.~C.,  {Hicks}
  B.~C.,  {Stewart} K.~P.,  {Cohen} A.~S.,    {Lane} W.~M.,  2004,
  Planet.~Space~Sci., 52, 1343

\bibitem[\protect\citeauthoryear{{Komatsu}, {Smith}, {Dunkley} \& {et
  al}}{{Komatsu} et~al.}{2010}]{Komatsu10}
{Komatsu} E.,  {Smith} K.~M.,  {Dunkley} J.,    {et al} 2010, preprint
  (astro-ph/1001.4538)

\bibitem[\protect\citeauthoryear{{Lidz}, {Zahn}, {McQuinn}, {Zaldarriaga} \&
  {Hernquist}}{{Lidz} et~al.}{2008}]{Lidz08}
{Lidz} A.,  {Zahn} O.,  {McQuinn} M.,  {Zaldarriaga} M.,    {Hernquist} L.,
  2008, ApJ, 680, 962

\bibitem[\protect\citeauthoryear{{Liu}, {Tegmark}, {Bowman}, {Hewitt} \&
  {Zaldarriaga}}{{Liu} et~al.}{2009}]{Liu09b}
{Liu} A.,  {Tegmark} M.,  {Bowman} J.,  {Hewitt} J.,    {Zaldarriaga} M.,
  2009, MNRAS, 398, 401

\bibitem[\protect\citeauthoryear{{Liu}, {Tegmark} \& {Zaldarriaga}}{{Liu}
  et~al.}{2009}]{Liu09a}
{Liu} A.,  {Tegmark} M.,    {Zaldarriaga} M.,  2009, MNRAS, 394, 1575

\bibitem[\protect\citeauthoryear{{Loeb} \& {Zaldarriaga}}{{Loeb} \&
  {Zaldarriaga}}{2004}]{Loeb04}
{Loeb} A.,  {Zaldarriaga} M.,  2004, Phys.~Rev.~Lett., 92, 211301

\bibitem[\protect\citeauthoryear{{Lonsdale}, {Cappallo}, {Morales}, {Briggs},
  {Benkevitch}, {Bowman}, {Bunton}, {Burns} \& {et al}}{{Lonsdale}
  et~al.}{2009}]{Lonsdale09}
{Lonsdale} C.~J. {et al.}, 2009, IEEE Proceedings, 97, 1497

\bibitem[\protect\citeauthoryear{{Madau}, {Meiksin} \& {Rees}}{{Madau}
  et~al.}{1997}]{Madau97}
{Madau} P.,  {Meiksin} A.,    {Rees} M.~J.,  1997, ApJ, 475, 429

\bibitem[\protect\citeauthoryear{{Mao}, {Tegmark}, {McQuinn}, {Zaldarriaga} \&
  {Zahn}}{{Mao} et~al.}{2008}]{Mao08}
{Mao} Y.,  {Tegmark} M.,  {McQuinn} M.,  {Zaldarriaga} M.,    {Zahn} O.,  2008,
  Phys.~Rev.~D, 78, 023529

\bibitem[\protect\citeauthoryear{{Masui}, {McDonald} \& {Pen}}{{Masui}
  et~al.}{2010}]{Masui10}
{Masui} K.~W.,  {McDonald} P.,    {Pen} U.,  2010, Phys.~Rev.~D, 81, 103527

\bibitem[\protect\citeauthoryear{{McQuinn}, {Lidz}, {Zahn}, {Dutta},
  {Hernquist} \& {Zaldarriaga}}{{McQuinn} et~al.}{2007}]{McQuinn07}
{McQuinn} M.,  {Lidz} A.,  {Zahn} O.,  {Dutta} S.,  {Hernquist} L.,
  {Zaldarriaga} M.,  2007, MNRAS, 377, 1043

\bibitem[\protect\citeauthoryear{{McQuinn}, {Zahn}, {Zaldarriaga}, {Hernquist}
  \& {Furlanetto}}{{McQuinn} et~al.}{2006}]{McQuinn06}
{McQuinn} M.,  {Zahn} O.,  {Zaldarriaga} M.,  {Hernquist} L.,    {Furlanetto}
  S.~R.,  2006, ApJ, 653, 815

\bibitem[\protect\citeauthoryear{{Mezzacappa}}{{Mezzacappa}}{2005}]{Mezzacappa%
05}
{Mezzacappa} A.,  2005, Annual Review of Nuclear and Particle Science, 55, 467

\bibitem[\protect\citeauthoryear{{Morales}}{{Morales}}{2005}]{Morales05}
{Morales} M.~F.,  2005, ApJ, 619, 678

\bibitem[\protect\citeauthoryear{{Morales}, {Bowman}, {Cappallo}, {Hewitt} \&
  {Lonsdale}}{{Morales} et~al.}{2006}]{Morales06MWA}
{Morales} M.~F.,  {Bowman} J.~D.,  {Cappallo} R.,  {Hewitt} J.~N.,
  {Lonsdale} C.~J.,  2006, New Astronomy Review, 50, 173

\bibitem[\protect\citeauthoryear{{Morales}, {Bowman} \& {Hewitt}}{{Morales}
  et~al.}{2006}]{Morales06}
{Morales} M.~F.,  {Bowman} J.~D.,    {Hewitt} J.~N.,  2006, ApJ, 648, 767

\bibitem[\protect\citeauthoryear{{Morales} \& {Hewitt}}{{Morales} \&
  {Hewitt}}{2004}]{Morales04}
{Morales} M.~F.,  {Hewitt} J.,  2004, ApJ, 615, 7

\bibitem[\protect\citeauthoryear{{Morales} \& {Wyithe}}{{Morales} \&
  {Wyithe}}{2010}]{Morales10}
{Morales} M.~F.,  {Wyithe} J.~S.~B.,  2010, ARA\&A, 48, 127

\bibitem[\protect\citeauthoryear{{Myers}, {Contaldi}, {Bond}, {Pen},
  {Pogosyan}, {Prunet}, {Sievers}, {Mason}, {Pearson}, {Readhead} \&
  {Shepherd}}{{Myers} et~al.}{2003}]{Myers03}
{Myers} S.~T.,  {Contaldi} C.~R.,  {Bond} J.~R.,  {Pen} U.,  {Pogosyan} D.,
  {Prunet} S.,  {Sievers} J.~L.,  {Mason} B.~S.,  {Pearson} T.~J.,  {Readhead}
  A.~C.~S.,    {Shepherd} M.~C.,  2003, ApJ, 591, 575

\bibitem[\protect\citeauthoryear{{Parsons}, {Backer}, {Foster}, {Wright},
  {Bradley}, {Gugliucci}, {Parashare}, {Benoit}, {Aguirre}, {Jacobs},
  {Carilli}, {Herne}, {Lynch}, {Manley} \& {Werthimer}}{{Parsons}
  et~al.}{2010}]{Parsons10}
{Parsons} A.~R.,  {Backer} D.~C.,  {Foster} G.~S.,  {Wright} M.~C.~H.,
  {Bradley} R.~F.,  {Gugliucci} N.~E.,  {Parashare} C.~R.,  {Benoit} E.~E.,
  {Aguirre} J.~E.,  {Jacobs} D.~C.,  {Carilli} C.~L.,  {Herne} D.,  {Lynch}
  M.~J.,  {Manley} J.~R.,    {Werthimer} D.~J.,  2010, AJ, 139, 1468

\bibitem[\protect\citeauthoryear{{Pearson} \& {Kus}}{{Pearson} \&
  {Kus}}{1978}]{Pearson78}
{Pearson} T.~J.,  {Kus} A.~J.,  1978, MNRAS, 182, 273

\bibitem[\protect\citeauthoryear{{Pen}, {Chang}, {Hirata}, {Peterson}, {Roy},
  {Gupta}, {Odegova} \& {Sigurdson}}{{Pen} et~al.}{2009}]{Pen09}
{Pen} U.,  {Chang} T.,  {Hirata} C.~M.,  {Peterson} J.~B.,  {Roy} J.,  {Gupta}
  Y.,  {Odegova} J.,    {Sigurdson} K.,  2009, MNRAS, 399, 181

\bibitem[\protect\citeauthoryear{{Peterson}, {Pen} \& {Wu}}{{Peterson}
  et~al.}{2004}]{Peterson04}
{Peterson} J.,  {Pen} U.,    {Wu} X.,  2004, preprint (astro-ph/0404083)

\bibitem[\protect\citeauthoryear{{Pritchard} \& {Loeb}}{{Pritchard} \&
  {Loeb}}{2010}]{Pritchard10}
{Pritchard} J.~R.,  {Loeb} A.,  2010, Phys.~Rev.~D, 82, 023006

\bibitem[\protect\citeauthoryear{{Purcell} \& {Field}}{{Purcell} \&
  {Field}}{1956}]{Purcell56}
{Purcell} E.~M.,  {Field} G.~B.,  1956, ApJ, 124, 542

\bibitem[\protect\citeauthoryear{{R{\"o}ttgering}, {Braun}, {Barthel}, {van
  Haarlem}, {Miley}, {Morganti}, {Snellen}, {Falcke}, \& {et al}}
  {{R{\"o}ttgering} et~al.}{2006}]{Rottgering06}
{R{\"o}ttgering} H.~J.~A. {et al.},  2006, preprint (astro-ph/0610596)

\bibitem[\protect\citeauthoryear{{Roy}, {Gupta}, {Pen}, Peterson, {Kudale} \&
  {Kodilkar}}{{Roy} et~al.}{2010}]{Roy10}
{Roy} J.,  {Gupta} Y.,  {Pen} U.,  Peterson J.~B.,  {Kudale} S.,    {Kodilkar}
  J.,  2010, Experimental Astronomy, 28, 25

\bibitem[\protect\citeauthoryear{{Salvaterra}, {Haardt} \&
  {Volonteri}}{{Salvaterra} et~al.}{2007}]{Salvaterra07}
{Salvaterra} R.,  {Haardt} F.,    {Volonteri} M.,  2007, MNRAS, 374, 761

\bibitem[\protect\citeauthoryear{{Scott} \& {Rees}}{{Scott} \&
  {Rees}}{1990}]{Scott90}
{Scott} D.,  {Rees} M.~J.,  1990, MNRAS, 247, 510

\bibitem[\protect\citeauthoryear{{Shapiro}, {Ahn}, {Alvarez}, {Iliev}, {Martel}
  \& {Ryu}}{{Shapiro} et~al.}{2006}]{Shapiro06}
{Shapiro} P.~R.,  {Ahn} K.,  {Alvarez} M.~A.,  {Iliev} I.~T.,  {Martel} H.,
  {Ryu} D.,  2006, ApJ, 646, 681

\bibitem[\protect\citeauthoryear{{Sunyaev} \& {Zeldovich}}{{Sunyaev} \&
  {Zeldovich}}{1975}]{Sunyaev75}
{Sunyaev} R.~A.,  {Zeldovich} I.~B.,  1975, MNRAS, 171, 375

\bibitem[\protect\citeauthoryear{{Thompson}, {Moran} \& {Swenson}
  Jr.}{{Thompson} et~al.}{2001}]{Thompson01}
{Thompson} A.~R.,  {Moran} J.~M.,    {Swenson} Jr. G.~W.,  2001,
  {Interferometry and Synthesis in Radio Astronomy, 2nd Edition}

\bibitem[\protect\citeauthoryear{{Trac} \& {Gnedin}}{{Trac} \&
  {Gnedin}}{2009}]{Trac09}
{Trac} H.,  {Gnedin} N.~Y.,  2009, preprint (astro-ph/0906.4348)

\bibitem[\protect\citeauthoryear{{Tseliakhovich} \& {Hirata}}{{Tseliakhovich}
  \& {Hirata}}{2010}]{Tseliakhovich10}
{Tseliakhovich} D.,  {Hirata} C.,  2010, preprint (astro-ph/1005.2416)

\bibitem[\protect\citeauthoryear{{Wang}, {Tegmark}, {Santos} \& {Knox}}{{Wang}
  et~al.}{2006}]{Wang06}
{Wang} X.,  {Tegmark} M.,  {Santos} M.~G.,    {Knox} L.,  2006, ApJ, 650, 529

\bibitem[\protect\citeauthoryear{{White}, {Carlstrom}, {Dragovan} \&
  {Holzapfel}}{{White} et~al.}{1999}]{White99}
{White} M.,  {Carlstrom} J.~E.,  {Dragovan} M.,    {Holzapfel} W.~L.,  1999,
  ApJ, 514, 12

\bibitem[\protect\citeauthoryear{{Willis}, {Oosterbaan} \& {de
  Ruiter}}{{Willis} et~al.}{1976}]{Willis76}
{Willis} A.~G.,  {Oosterbaan} C.~E.,    {de Ruiter} H.~R.,  1976, A\&AS, 25,
  453

\bibitem[\protect\citeauthoryear{{Willott}, {Delorme}, {Omont}, {Bergeron},
  {Delfosse}, {Forveille}, {Albert}, {Reyl{\'e}}, {Hill}, {Gully-Santiago},
  {Vinten}, {Crampton}, {Hutchings}, {Schade}, {Simard}, {Sawicki}, {Beelen} \&
  {Cox}}{{Willott} et~al.}{2007}]{Willott07}
{Willott} C.~J.,  {Delorme} P.,  {Omont} A.,  {Bergeron} J.,  {Delfosse} X.,
  {Forveille} T.,  {Albert} L.,  {Reyl{\'e}} C.,  {Hill} G.~J.,
  {Gully-Santiago} M.,  {Vinten} P.,  {Crampton} D.,  {Hutchings} J.~B.,
  {Schade} D.,  {Simard} L.,  {Sawicki} M.,  {Beelen} A.,    {Cox} P.,  2007,
  AJ, 134, 2435

\bibitem[\protect\citeauthoryear{{Willott}, {Rawlings} \& {Blundell}}{{Willott}
  et~al.}{2001}]{Willott01}
{Willott} C.~J.,  {Rawlings} S.,    {Blundell} K.~M.,  2001, MNRAS, 324, 1

\bibitem[\protect\citeauthoryear{{Wouthuysen}}{{Wouthuysen}}{1952}]{Wouthuysen%
52}
{Wouthuysen} S.~A.,  1952, AJ, 57, 31

\bibitem[\protect\citeauthoryear{{Zahn}, {Lidz}, {McQuinn} \& {Dutta}}{{Zahn}
  et~al.}{2007}]{Zahn07}
{Zahn} O.,  {Lidz} A.,  {McQuinn} M.,    {Dutta} S.~a.,  2007, ApJ, 654, 12

\bibitem[\protect\citeauthoryear{{Zaldarriaga}, {Furlanetto} \&
  {Hernquist}}{{Zaldarriaga} et~al.}{2004}]{Zaldarriaga04}
{Zaldarriaga} M.,  {Furlanetto} S.~R.,    {Hernquist} L.,  2004, ApJ, 608, 622

\bibitem[\protect\citeauthoryear{{Zaroubi} \& {Silk}}{{Zaroubi} \&
  {Silk}}{2005}]{Zaroubi05}
{Zaroubi} S.,  {Silk} J.,  2005, MNRAS, 360, L64

\end{thebibliography}

\end{document}